\documentclass[]{raa}            % referee version: for submission
\usepackage{graphicx,times}
\usepackage{natbib}

\providecommand{\bm}[1]{\mbox{\boldmath$#1$\unboldmath}}

\begin{document}

   \title{Dust growth in protoplanetary disks -- a comprehensive
   experimental/theoretical approach}

 \volnopage{ {\bf 20xx} Vol.\ {\bf xx} No. {\bf xx}, 000--000}
   \setcounter{page}{1}

   \author{J. Blum
      \inst{}
}

   \institute{Institut f\"ur Geophysik und extraterrestrische Physik,
   University of Braunschweig, Mendelssohnstr. 3,
   38106 Braunschweig, Germany; {\it j.blum@tu-braunschweig.de}\\
\vs \no
   {\small Received [year] [month] [day]; accepted [year] [month] [day] }
}

\abstract{More than a decade of dedicated experimental work on the collisional physics of
protoplanetary dust has brought us to a point at which the growth of dust aggregates can -- for the
first time -- be self-consistently and reliably modelled. In this article, the emergent collision
model for protoplanetery dust aggregates \citep{güttleretal2010} as well as the numerical model for
the evolution of dust aggregates in protoplanetary disks \citep{zsometal2010} are reviewed. It
turns out that, after a brief period of rapid collisional growth of fluffy dust aggregates to sizes
of a few centimeters, the protoplanetary dust particles are subject to bouncing collisions, in
which their porosity is considerably decreased. The model results also show that low-velocity
fragmentation can reduce the final mass of the dust aggregates but that it does not trigger a new
growth mode as discussed previously. According to the current stage of our model, the direct
formation of kilometer-sized planetesimals by collisional sticking seems impossible so that
collective effects, such as the streaming instability and the gravitational instability in
dust-enhanced regions of the protoplanetary disk, are the best candidates for the processes leading
to planetesimals.  }

   \authorrunning{J. Blum}
   \titlerunning{Dust growth in protoplanetary disks --
   a comprehensive experimental/theoretical approach}
   \maketitle

%% Note: In the following text body of your manuscript, please note several differences from
%%       other major journals:
%% (1) \subsection{Please Capitalize the First Letter of Each Notional Word in Subsection Title}
%% (2) Please Capitalize the First Letter of Each Notional Word in all tables' captions

%%\subsection{Please Capitalize the First Letter of Each Notional
%%Word in Subsection Title}

%%\subsubsection{This is a third-level section --- subsubsection}
%%Some applications of the routines are given in Table~\ref{Table
%%1}.

%
%________________________________________________ sections below
%
\section{Introduction and Motivation}           %% first-level sections will be auto-capitalized
\label{sect:introduction}

The `standard' paradigm of planet formation comprises a two-stage process: (1) When the particles
are small, i.e. when we speak of `dust', growth is supposed to be by {\it coagulation}, i.e. dust
particles stick to one another due to non-gravitational forces, e.g. surface adhesion (van der
Waals force). (2) For much larger particles, i.e. for `planetesimals', the further growth is based
on {\it accretion}, i.e. mutual gravitational attraction of the colliding bodies.

As the gravitational potential of an individual body increases
with its mass, stage (2) requires planetesimal sizes of at least
$\sim$ 1 km, i.e. escape velocities of the order of $1~\rm
m~s^{-1}$, before accretion in mutual collisions becomes
effective. Thus, we are faced with the problem to explain dust
growth in protoplanetary disks (PPDs) due to `sticking collisions'
for a range of dust masses spanning 27 orders of magnitude, i.e.
dust sizes ranging from $\sim 1 \rm ~\mu m$ to $\sim 1 \rm ~km$.

To better assess the scenario of dust growth in PPDs, we will first look at the prerequisites for
collisional evolution, i.e. the causes of mutual collisions among the dust grains. Current models
of PPDs (see, e.g., \citet{dullemondetal2007}) favor geometrically thin (but optically thick)
flared disk structures with the gas pressure and temperature decreasing radially outward and a
modest degree of MRI-driven turbulence in the gas. In such a scenario, the embedded dust particles
undergo a variety of random and systematic motions, which lead to frequent collisions among them
\citep{weidenschilling1977}. Very small particles (sizes $\stackrel{<}{\sim} 100 \rm \mu m$) are
mostly affected by Brownian motion, which leads to collisions at extremely low velocities, i.e. $v
\stackrel{<}{\sim} 10^{-3}~\rm m~s^{-1}$. Larger grains are subject to systematic motion with
respect to the nebular gas: particles outside the midplane of the PPD sediment towards the
midplane, due to the vertical component of the gravitational attraction of the central star; in
addition to that, dust particles spiral radially inward, owing to their friction with the slower
rotating (pressure-supported) gas disk. Both drift velocities increase with increasing
mass-to-surface ratio of the dust particles so that large dust aggregates catch up with small ones.
Turbulence in the gas also causes dust particles to collide with one another, also for dust
aggregates with identical masses. \citet{weidenschillingcuzzi1993} derived collision velocities for
all combinations of dust sizes, starting from small grains with sizes of $1 ~\rm \mu m$ all the way
to the smallest planetesimals. Besides the above-mentioned regime, where Brownian motion dominates
the collisional evolution of the dust grains and in which the collision velocity {\it decreases}
with the mass of the aggregates, the collision velocity typically {\it increases} with increasing
size of the dust grains up to meter-sized bodies, after which collision speeds stay rather constant
at values around $v \approx 50 ~\rm m~s^{-1}$. An improved model of turbulence-induced collision
velocities was published by \citet{ormelcuzzi2007} who give closed-form solutions for all
particle-size combinations. It should be mentioned that all relative velocities (with the exception
of those caused by Brownian motion) are a function of gas density and strength of turbulence,
usually characterized by an $\alpha$ value.

It will be shown later (see Sect. \ref{sect:model}) that a critical velocity in the collisional
evolution is at $v \approx 1 ~\rm m~s^{-1}$, above which dust aggregates tend to fragment in
collisions. In the minimum-mass solar nebula (MMSN) model with $\alpha \approx 10^{-4}$, assumed by
\citet{weidenschillingcuzzi1993}, this collision velocity is reached for cm-sized dust particles;
lower turbulence strength of $\alpha = 10^{-5}$ increases the size at which the fragmentation
velocity is reached to decimeters (see \citet{weidlingetal2009}). Other PPD models, having
different gas pressures, gas temperatures and pressure gradients, exhibit a similar velocity
systematics but can reach the critical velocity for fragmentation for very different dust-aggregate
sizes (see, e.g., \citet{weidlingetal2009,zsometal2010}).

Astronomical observations of PPDs at various wavelengths yield strong indications for grain growth
\citep{nattaetal2007}. Unfortunately, aggregate sizes in excess of a few mm cannot be detected due
to their inefficient thermal emission. However, the detection of mm-sized particles in PPDs is a
clear evidence of grain growth in such disks. Remarkably, mm-sized particles are also preserved
from our own PPD, the so-called solar nebula: in primitive meteorites a predominant part of the
mass is found in {\it chondrules}, mm-sized spherules with ages (determined by radio-isotope
dating) placing their formation within the first few million years of the solar nebula. Chondrules
were molten by an unknown process and solidified within a short time so that they survived the
process of the formation of their parent bodies. Chondrules provide strong evidence that (at least
at one location) in the solar nebula, (at least) mm-sized dust aggregates were present.

It is the objective of this article to unveil the growth
processes of protoplanetary dust from an experimental as well as
from a numerical point of view. We will see which physical
interactions dust aggregates undergo in mutual low-velocity
collisions and to which growth timescales, mass distributions and
dust-aggregate structures this leads. As it will turn out, the
wealth of laboratory investigations of the past decade will
severely challenge the planetesimal-formation scenario outlined
above, particularly by showing that dust aggregates tend to bounce
and fragment, rather than stick to one another, in collisions at
velocities $v \stackrel{>}{\sim} 1 ~\rm m~s^{-1}$. It will become
clear that collisional sticking alone cannot form km-sized
planetesimals.

\section{Collision and aggregation experiments with dust particles}
\label{sect:experiments}

It is an experimental challenge to approach the problem of dust growth in PPDs. Experimental parameters, such as dust-particle sizes, dust materials (oxides, metals, silicates, organic material, ices), collision velocities as a function of dust-aggregate size, dust-aggregate morphologies (fractal, porous, compact), the gaseous environment (temperature and pressure) and the charging state of the dust particles should match those of the dust in PPDs as closely as possible. It is evident from this list of parameters that one single experiment can never fulfil this. Thus, we chose the approach to piecewise match the dust-growth scenario in PPDs, starting from small dust and low velocities (e.g. Brownian motion) in a more or less self-consistent way all the way to larger aggregates with realistic morphologies and higher collision speeds. The experimental approach is ideally paralleled with numerical simulations of the dust growth, taking into account the experimental results as well as physical models for the static and dynamical interactions between dust particles.

\begin{figure}[h!!!]
   \centering
   \includegraphics[width=9.0cm, angle=0]{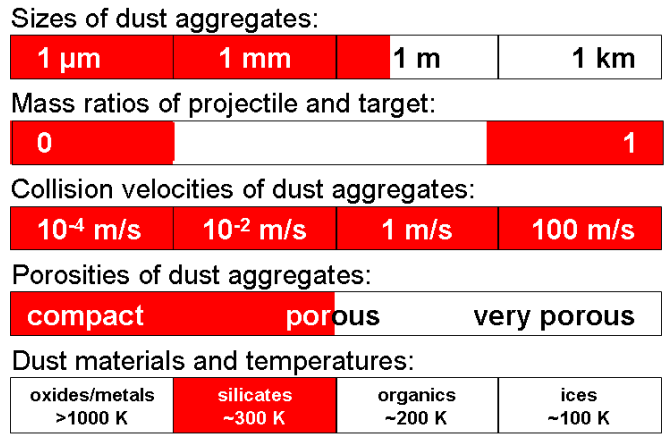}

   \begin{minipage}[]{85mm}

   \caption{\label{fig:completeness}Schematic illustration of the parameter
   space of dust-aggregate collision experiments simulating protoplanetary dust    evolution. The red-shaded areas have been covered by experiments, whereas the white regions are (yet) unexplored.} \end{minipage}
   \end{figure}

Due to the overwhelming size of the parameter space to be covered by this approach, first results
are only available for a limited set of parameters (see \citet{güttleretal2010} and
\citet{zsometal2010} for details). Fig. \ref{fig:completeness} schematically shows where we
currently stand in terms of completeness. (1) The sizes of protoplanetary dust range from initially
$\rm \sim 1 ~ \mu m$ to planetesimal dimensions of $\stackrel{>}{\sim} 1$ km, above which gravity
becomes the dominant effect in collisions. It is obvious that this size range can never be covered
in laboratory experiments. However, for all dust-aggregate sizes treatable in the laboratory ($\rm
\sim 1 ~ \mu m \ldots ~ 0.1~m$), experiments have been performed. (2) The mass ratio between the two
dust aggregates that collide in the protoplanetary nebula is also a potential parameter to cover in
laboratory experiments. It is not difficult to imagine that an impact of a micrometer-sized dust
grain into a cm-sized dust aggregate has a different outcome compared to a collision between two
cm-sized aggregates at the same velocity. Currently, laboratory collision experiments with rather
similar-sized dust aggregates as well as those with aggregates of very dissimilar sizes have been
performed. However, we are yet far away from a complete coverage of this parameter. (3) The
completeness of the experimental investigations is best for the collision velocity. The
experimental velocities achieved so far reach from $\sim \rm 10^{-4}~m~s^{-1}$ for Brownian motion
to $\sim \rm 100~m~s^{-1}$ for ballistic impacts. (4) It is also not hard to imagine that the
collision velocity influences the morphology of the growing dust aggregates. Extremely low
velocities will result in very open-structured, porous dust aggregates, whereas collisions at
higher velocities will lead to rather compact dusty bodies. Thus, experiments with arbitrary
porosities of the aggregates are desirable. The coverage of the potential porosities is, however,
still rather poor. While experiments with compact dust aggregates are easy to perform, there is
only a limited number of collision experiments with very porous dust aggregates available. (5)
Finally, the question of the dust materials to be used in analog experiments needs to be addressed:
from the composition of the major bodies in the Solar System and from calculations of the
condensation sequences in protoplanetary disks, four major groups of materials are expected that
each dominate their own region within the PPD \citep{lewis1972,grossman1972}: in the inner region,
where the temperatures are highest, the dust composition is dominated by refractory materials, like
oxides or metals; further out, the less refractory silicates condense, followed by the condensation
of organic materials and ices in the outer regions of the PPD. Most of the laboratory experiments
on protoplanetary dust agglomeration have so far been performed with silicates and only very few
dealt with the other protoplanetary dust materials.

The most complete set of experimental investigations on the collision and growth behavior of
protoplanetary dust has been using $\rm SiO_2$ as the dust material, with particle sizes around
$\rm 1 ~\mu m$ (see \citet{blumwurm2008} for details).  A large subset of these experiments used
idealized dust particles, consisting of monodisperse, spherical $\rm SiO_2$ grains of $0.75 ~ \rm
\mu m$ radius \citep{güttleretal2010}. Before we review these experiments and the modeling approach
for {\it dust aggregates}, we have to address two basic questions concerning {\it individual dust
particles}:

\begin{enumerate}

\item Why do {\it individual dust particles} adhere to one another?\\ The formation of dust
aggregates in the terrestrial-planet region of PPDs most likely happens in a charge-free
environment, the so-called MRI-dead zone \citep{gammie1996,terquem2008}. However, triboelectric
charging in mutual collisions may provide a source for enhanced stickiness of small aggregates
\citep{poppeetal2000b,marshalletal2005} but is unlikely to be responsible for the overall growth of
protoplanetary dust. Thus, we can assume that dust particles are basically uncharged so that
Coulomb forces cannot be responsible for the `stickiness' of the dust. In the absence of free
charge carriers, dust particles do only interact through van der Waals forces, i.e. stochastic
dipole interactions, leading to extremely small and short-ranged cohesion forces.
\citet{heimetal1999} used an atomic force microscope, to which they glued individual spherical $\rm
SiO_2$ particles with different diameters, and measured the separation forces between these
particles and other spherical $\rm SiO_2$ particles, which were glued on a flat substrate. Thus,
they were able to confirm the predicted proportionality between binding force and particle radius
\citep{johnsonetal1971,derjaguinetal1975}. \citet{heimetal1999} derived typical binding forces of
$\sim 10^{-7}$~N for micrometer-sized particles.

\item At which impact velocities do {\it individual dust particles} stick upon a
collision?\\\citet{poppeetal2000} performed impact experiments of the same type of
micrometer-sized, spherical $\rm SiO_2$ particles onto flat $\rm SiO_2$ substrates. They used a
cogwheel dust deagglomerator, which is able to separate the individual dust particles of a powder
sample and to accelerate the dust grains in a jet-like fashion \citep{poppeetal1997}. The impacts
of the dust grains onto the targets were observed with a high-resolution long-distance microscope
with a pulsed-laser illumination so that the particle trajectories before and after impact could be
determined. \citet{poppeetal2000} found a rather sharp transition from sticking (for $v < v_{\rm
c}$) to bouncing (for $v > v_{\rm c}$) at a velocity of $v_{\rm c} \approx 1 ~\rm m~s^{-1}$. This
threshold velocity increased with decreasing particle size.

\end{enumerate}

\begin{figure}[h!!!]
   \centering
   \includegraphics[width=9.0cm, angle=0]{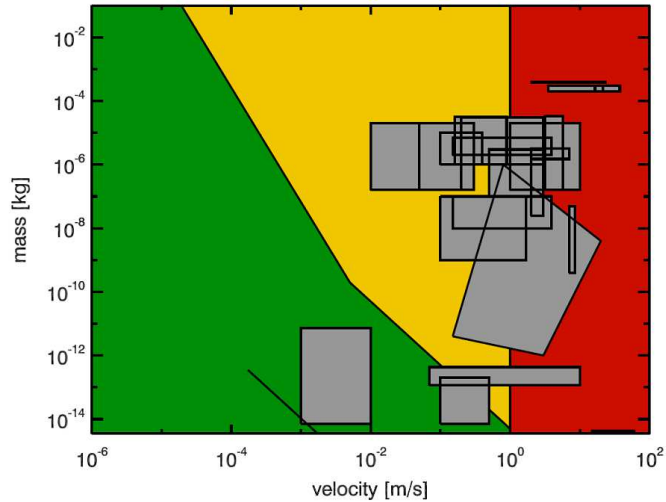}

   \begin{minipage}[]{85mm}

   \caption{\label{fig:parameterspace}Dust-aggregate masses and collision velocities of the existing dust-growth experiments. Overplotted are the regions in which sticking (green), bouncing (yellow), and fragmentation (red) are expected. Data taken from \citet{güttleretal2010} with augmentation by our latest laboratory experiments.} \end{minipage}
   \end{figure}

Knowing the adhesion and collision properties of the individual $\rm SiO_2$ monomer, we can now
concentrate on the growth and collision behavior of dust aggregates. \citet{güttleretal2010}
reviewed a set of 19 different experiments, which all deal with the evolution of protoplanetary
dust aggregates. On top of that, a few new experiments were launched in our laboratory in the past
months. Fig. \ref{fig:parameterspace} shows the masses of the dust aggregates and the collision
velocities of these experiments. In the following, we will consider some of these experiments in
more detail. One should bear in mind that -- although the experimental coverage of dust-aggregate
masses and collision velocities is quite satisfactory (see Fig. \ref{fig:completeness}) -- the
two-dimensional parameter space shows considerable regions that are relevant to protoplanetary dust
growth and are not covered by experiments(see Fig. \ref{fig:parameterspace}).

The experiments dealing with the initial stage of dust agglomeration in PPDs used homogeneous
dispersions of individual micrometer-sized dust particles and observed the agglomeration in
low-velocity collisions. From the work of \citet{poppeetal2000} it is clear that the collision
velocities have to be below $1\rm ~m~s^{-1}$ so that the sticking probability is close to unity. As
the sources for the collision of small particles in PPDs are mainly Brownian motion, relative drift
motions, and gas turbulence, experiments concentrated on these effects to observe the onset of
agglomeration \citep{blumetal2000,krauseblum2004,wurmblum1998,blumetal1998}. All these experiments
found consistently and in agreement with numerical simulations
\citep{kempfetal1999,paszundominik2006} that dust agglomeration in PPDs starts with a {\it fractal}
growth regime, in which the colliding dust particles stick at the first contact (hit-and-stick
growth). The main characteristics of this growth regime are (1) the fractal nature of the growing
dust aggregates with a mass ($m$) - size ($s$) relation $m \propto s^D$, with $D < 2$ being their
fractal dimension, (2) a quasi-monodisperse mass distribution at any given time $t$, and (3) a
power-law temporal growth of the mean mass $<m> \sim t^2$ (see \citet{blum2006} for details). For
aggregate sizes exceeding $\sim 100 ~\rm \mu m$, Brownian motion becomes slower than the drift
motion so that the collision energy increases with increasing aggregate masses. If this energy
exceeds a threshold above which frictional forces are no longer sufficient to allow for a
hit-and-stick behavior, the dust aggregates are compacted upon collision
\citep{dominiktielens1997,blumwurm2000,wadaetal2007,wadaetal2008,wadaetal2009}. Thus, it is clear
that the `compactness' or `fluffiness' of a dust aggregate is an important parameter in the
evolution of protoplanetary dust. We describe this `compactness' or `fluffiness' either by the
porosity of the aggregate (see Fig. \ref{fig:completeness}) or by the enlargement parameter $\Psi =
V/V_{\rm c}$, where $V$ and $V_{\rm c}$ describe the actual volume and the compact volume of a dust
aggregate. The latter is related to the mass of the aggregate through $V_{\rm c} = m/\rho_0$, with
$\rho_0$ being the mass density of the monomer grains. It can easily be imagined that it becomes
increasingly difficult to perform laboratory experiments with higher and higher porosities.
`Natural' dust aggregates, as they occur in powder samples have porosities of typically 60\% (i.e.
enlargement parameters of $\Psi = 2.5$). Denser aggregates (down to enlargement parameters of $\Psi
= 1.7$) can be manufactured by compressing a dust sample; looser (but still coherent) aggregates
can be made by the process of random ballistic deposition \citep{blumschraepler2004}, in which
individual micrometer-sized dust particles are added one-by-one in a hit-and-stick fashion to grow
macroscopic bodies. Random ballistic deposition leads to porosities of 85\% (i.e. to enlargement
parameters of $\Psi = 6.7$) for monodisperse spherical dust particles. Recipes for the
manufacturing of large dust aggregates with higher porosities have not been published, although the
models for the enlargement parameter of protoplanetary dust aggregates predict values of to $\Psi =
15-35$ \citep[see also Fig. \ref{fig:result}b in Sect. \ref{sect:results}]{zsometal2010}.

\begin{figure}[h!!!]
   \centering
   \includegraphics[width=9.0cm, angle=0]{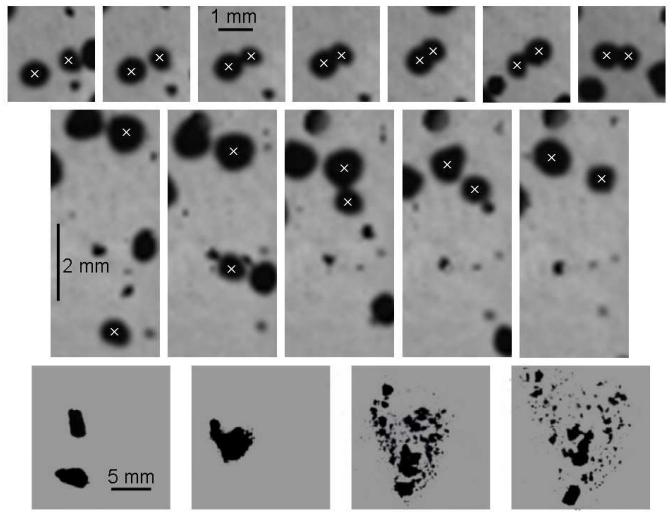}

   \begin{minipage}[]{85mm}

   \caption{\label{fig:sbf}Examples of laboratory collision experiments between
   fluffy $\rm SiO_2$ aggregates, which resulted in sticking (top), bouncing (center),
   and fragmentation (bottom). The collision velocities were 0.004 $\rm m~s^{-1}$
   (sticking), 0.16 $\rm m~s^{-1}$ (bouncing), and 5.1 $\rm m~s^{-1}$ (fragmentation),
   respectively. All experiments were performed under microgravity conditions.
   For clarity, the colliding dust aggregates in the sticking and bouncing case
   are marked by an x. The time interval between the first five images in the
   sticking case is 54 ms, the last three images have time intervals of 850 ms
   between them to proof the durability of the adhesion; the time interval
   between the images in the bouncing case is 14 ms; the time interval between the
   first two images and the last two images in the case of fragmentation
   is 1 ms, respectively.} \end{minipage}

   \end{figure}

Experiments on the collisional behavior of dust aggregates with rather high porosities ($\Psi =
6.7$, manufactured by the random ballistic deposition process) were performed by
\citet{langkowskietal2008} and \citet{salteretal2009}. In combination with the experiments using
fractal dust aggregates \citep{blumetal2000,krauseblum2004,wurmblum1998,blumetal1998} and those
with dust aggregates having low porosities (see \citet{güttleretal2010} for details), the emergent
picture is that three different collisional outcomes can be distinguished: (1) sticking, in which
the mass of at least the larger of the two colliding dust aggregates is increased, (2) bouncing, in
which the mass of the interacting bodies is basically unchanged, and (3) fragmentation, in which
the mass of the impacting dust aggregates is reduced. Fig. \ref{fig:sbf} shows experimental
examples for sticking, bouncing, and fragmentation. Counter-intuitively, fragmentation {\it can} --
under certain circumstances -- also lead to a mass increase of one (generally the larger) of the
dust aggregates \citep{wurmetal2005}: if a (smaller) projectile of arbitrary porosity impacts a
(larger) target aggregate, which is compact (i.e. having enlargement parameters of $\Psi
\stackrel{<}{\sim} 2$), at velocities exceeding the fragmentation threshold (i.e. for $v > \rm
1~m~s^{-1}$), part of the projectile aggregate (typically a few ten percent of its mass) sticks to
the target, thus enhancing the mass of the larger collision partner.

A thorough look at the published protoplanetary-dust experiments reveals that yet another parameter
is required to describe the collisional physics of fluffy dust aggregates, i.e. the {\it mass
ratio} of the colliding dust aggregates. Experiments by \citet{langkowskietal2008}, who observed
the impacts of mm-sized high-porosity dust aggregates into cm-sized high-porosity target aggregates
at various velocities, showed that for colliding dust aggregates with the same porosity but very
different sizes the smaller (projectile) aggregate gets buried in the larger (target) aggregate
for $v \stackrel{>}{\sim} 1 ~\rm m~s^{-1}$. On the other hand, if, for the identical dust-aggregate
morphologies and porosities, the target size is reduced to the projectile size, the experiments
show that both aggregates fragment for collision velocities $v \stackrel{>}{\sim} 1 ~\rm m~s^{-1}$
(see Fig. \ref{fig:sbf}, bottom). This means that there is growth for projectile-target collisions,
whereas the opposite (i.e. fragmentation) happens if both aggregates are similar in size. For a
realistic collision model of protoplanetary dust aggregates it is therefore mandatory to take into
account four parameters:
\begin{enumerate}
  \item the collision velocity,
  \item the mass of the projectile aggregate, i.e. the aggregate with the smaller mass,
  \item the mass ratio between the two colliding aggregates,
  \item the porosity of the aggregates, i.e. their enlargement parameter.
\end{enumerate}
Thus, for a deeper understanding of the evolution of protoplanetary dust aggregates, a full
coverage of this four-dimensional parameter space (for a variety of realistic dust materials) is
required. In the following Section, we will describe the main characteristics of our collision
model, which takes into account this multi-parameter space. For details, refer to
\citet{güttleretal2010}.

\section{A new dust-aggregate collision model}
\label{sect:model} To be able to predict the temporal evolution of protoplanetary dust aggregates,
a physical model on the collision behavior of arbitrary dusty bodies is mandatory. As we have seen
above, a full four-dimensional treatment of the parameter space (mass, mass ratio, porosity, and
collision velocity of the dust aggregates) is, however, not yet possible because the coverage of
the mass ratio and enlargement parameter is far from being complete. Thus, we decided to treat
these two parameters in a binary way in our model \citep{güttleretal2010}. We describe dust
aggregates either as `porous' (p) or as `compact' (c). Without better knowledge, we set the
threshold between the two regimes at an enlargement parameter of $\Psi = 2.5$ (i.e. a porosity of
60\%), i.e. all aggregates with $\Psi > 2.5$ are `porous' and those with $\Psi < 2.5$ are
considered `compact'. In a similar way, the mass ratio of the colliding aggregate pair is
parameterized: for mass ratios between the more and less massive dust aggregate $r < 100$, we
consider both aggregates to be of equal size; for $r > 100$, we treat the collision as an impact of
a projectile into an infinitely large target aggregate. Under the assumption of similar porosity of
the two colliding dust aggregates, the threshold mass ratio of $r = 100$ corresponds to a size
ratio of 4.6. With this simplification, the following eight collision types are possible and are
independently treated in our collision model: `pp', `pP', `pc', `pC', `cp', `cP', `cc', and `cC'.
Here, two small letters characterize a collision with $r < 100$, whereas a combination of a small
and a capital letter depicts a projectile-target collision. Please mind that the `pc' combination
differs from `cp' in such a way that in `pc' the (slightly) less-massive of the two colliding
aggregates is porous, whereas in the `cp' case it is compact.

\begin{figure}[h!!!]
   \centering
   \includegraphics[width=15.0cm, angle=0]{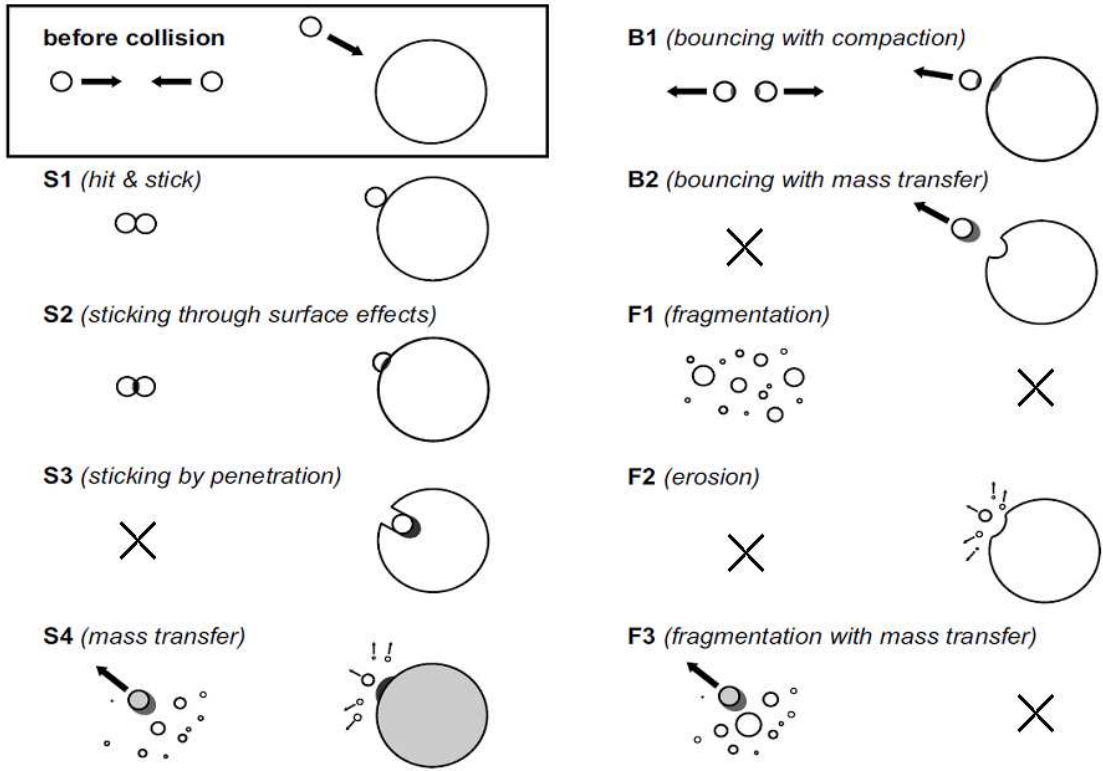}

   %\begin{minipage}[]{85mm}

   \caption{\label{fig:pictograms}Pictorial representations of possible outcomes in protoplanetary dust-aggregate collisions (taken from \citet{güttleretal2010}). On the respective top parts, collisions between similar-sized aggregates $r < 100$ are depicted, whereas the bottom parts show collisions of the projectile-target type ($r > 100$). Gray shading denotes that this particular type of collision happens for compact particles ($Psi < 2.5$) only. The collisional outcomes distinguish between sticking (S1-S4), bouncing (B1-B2), and fragmenting (F1-F3) collisions (see Sect. \ref{sect:model} for details).}

   %\end{minipage}
   \end{figure}

The laboratory experiments have shown that a simple distinction between sticking, bouncing, and fragmentation is insufficient to describe the full suite of possible outcomes of inter-aggregate collisions. Taking into account all cases seen in laboratory experiments, our collision model was generalized and distinguishes between four different types of sticking (S1-S4), in which the more massive of the collision partners gains mass, two types of bouncing (B1-B2), and three types of fragmentation (S1-S3), in which the more massive aggregates loses mass. Fig. \ref{fig:pictograms} shows pictorial representations of these nine collision types.

We now briefly describe the main physical effects leading to the various collision types. The interested reader is referred to \citet{güttleretal2010} for a detailed description and the complete set of formulae.
\begin{itemize}
  \item S1: hit-and-stick collisions.\\As described above, the two colliding dust aggregates stick at their first point of contact.   This growth type leads to fractal dust aggregates and is limited by the threshold velocity for compaction as an upper velocity limit for S1. The experimental evidence for S1 can be found in \citet{blumetal2000,krauseblum2004,wurmblum1998,blumetal1998}.
  \item S2: sticking through surface effects.\\For velocities above the S1 limit, dust aggregates can in principle still stick upon a collision if they are able to dissipate sufficient energy and if the contact surface is large enough. Collisional compaction is a process that does both, i.e. it increases the contact surface and is mainly responsible for the dissipation of kinetic energy. Using a Hertzian model for the deformation of fluffy dust aggregates, which was `calibrated' with the experimental findings, we derived upper limits for dust-aggregate sticking. Fig. \ref{fig:sbf} (top) shows an example for S2.
  \item S3: sticking by penetration.\\If the collision velocities are above the S2 limit, dust aggregates can still gain mass. Process S3 is one method by which this is possible: in the case of projectile-target  collisions and porous targets, the projectile aggregate does not fragment upon impact but is -- above a threshold velocity -- embedded in the target aggregate, leading to a mass gain of the target \citep{langkowskietal2008}.
  \item S4: sticking through mass transfer.\\If, on the other hand, the  larger of the two projectiles is compact and the less massive aggregate is porous, and if the collision velocity is above the S2 limit, the porous aggregate is likely to fragment in the collision. It was observed in experiments \citep{wurmetal2005} that this type of collision results in the attachment of part of the fluffy aggregate's mass to the compact particle.
  \item B1: bouncing with compaction.\\In those cases, in which the  collision velocity is above the S2 threshold and below the threshold   for fragmentation or sticking by penetration, the aggregates bounce off. Previous experiments had shown that -- even in the case of bouncing -- a substantial part of the kinetic energy is dissipated, typically between 80\% and 95\% \citep{blummuench1993}. Further experiments by \citet{weidlingetal2009} showed that the energy is used to restructure the aggregates close to the surface, which leads to local compaction. Thus, repetitive bouncing is a means to convert a `porous' into a `compact' dust aggregate. When, after repeated compaction processes, the dust aggregates are sufficiently dense, they might also fragment at velocities below the threshold of process F1 with a small likelihood.
  \item B2: bouncing with mass transfer.\\Bouncing can also be accompanied  by a transfer of mass for projectile-target collision between porous  dust aggregates. \citet{langkowskietal2008} found that this mass transfer always leads to a mass increase of the (smaller) projectile and never of the (larger) target. In B2 collisions, the projectile aggregates typically doubles its mass.
  \item F1: fragmentation.\\At sufficiently high velocities, the internal  strength of the dust aggregates is not sufficient to withstand the  impact stresses so that the dust aggregates break into smaller fragments. The fragment masses follow a velocity-independent power law, with the largest fragment being a function of impact speed \citep{blummuench1993,güttleretal2010}. For the micrometer-sized spherical $\rm SiO_2$ particles used in most of our experiments, the fragmentation boundary is at $v_{\rm c} \approx 1 ~\rm m~s^{-1}$, remarkably close to the threshold velocity for sticking of the monomer grains (see Sect. \ref{sect:experiments}).
  \item F2: erosion.\\Very small dust particles (e.g. monomer grains) can  erode the surface of a much larger dust aggregate. This process initially leads to the mass loss of the porous target aggregate but also compacts it such that the target aggregates gets passivated against erosion up to velocities of $> 10 ~\rm m~s^{-1}$ \citep{schraeplerblum2010}.
  \item F3: fragmentation with mass transfer.\\Very similar to process F1 and S4, collisions between similar-sized aggregates lead to the   fragmentation of the porous aggregates. If the (slightly) larger of  the two aggregates is porous and the smaller aggregate is compact, part of the mass of the porous aggregate is transferred to the compact aggregate so that the larger aggregate loses mass and the smaller one gains mass \citep{güttleretal2010}. We count this as fragmentation, as the more massive object is fragmented.
\end{itemize}

\begin{figure}[h!!!]
   \centering
   \includegraphics[width=15.0cm, angle=0]{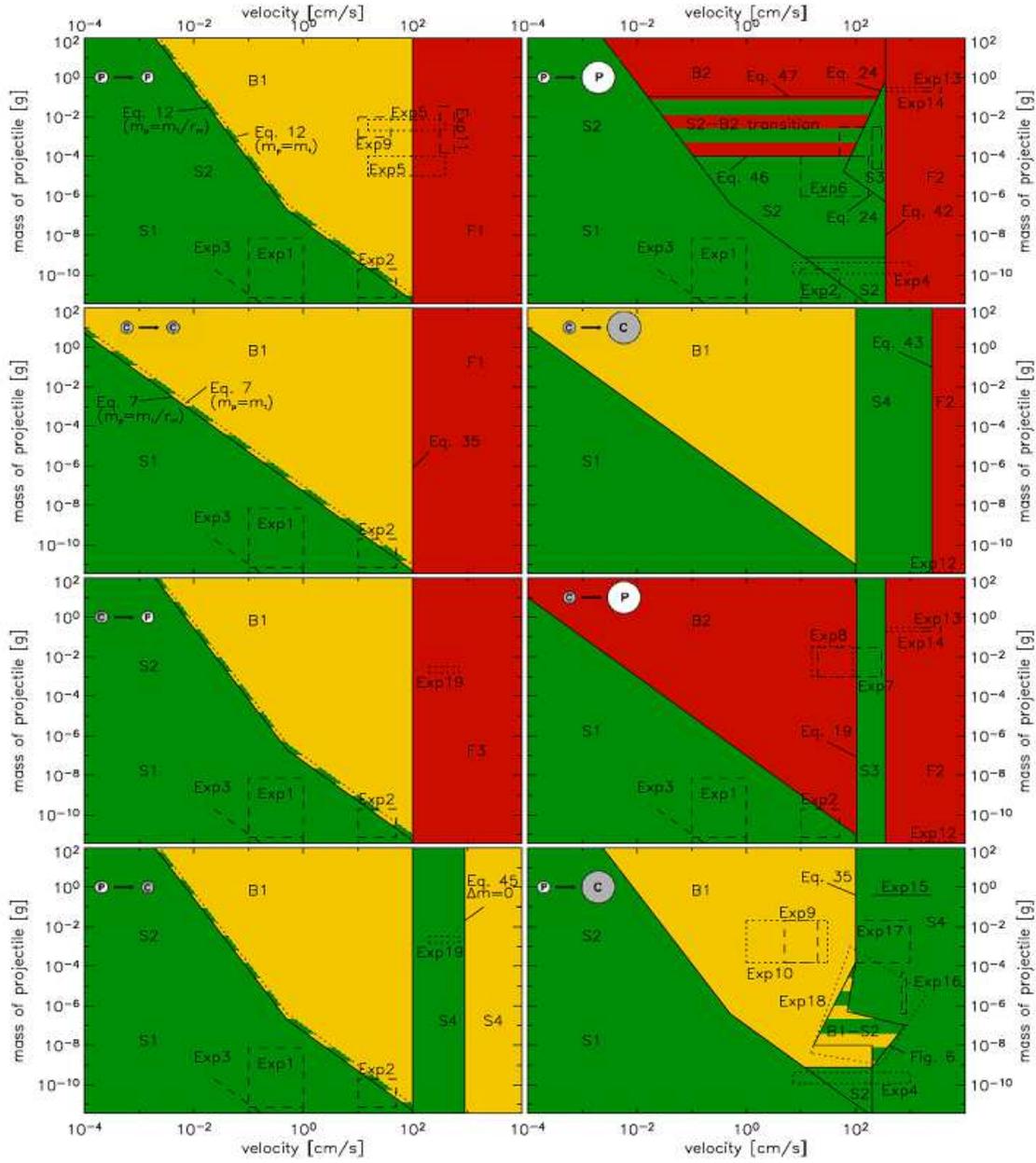}

   %\begin{minipage}[]{85mm}

   \caption{\label{fig:model}The collision model for protoplanetary
   dust aggregates in the mass-velocity parameter space for the eight
   combinations of aggregate mass ratios and porosities (adapted from
   \citet{güttleretal2010}).}
   %\end{minipage}
   \end{figure}

It must be emphasized again that the experimental coverage of the parameter space is quite sparse
so that a complete collision model has to rely on extrapolations from and interpolations between
experimental data points. On top of that, the collision model requires quite a large number of
physical parameters as input values (see Table 2 in \citet{güttleretal2010}). We carefully measured
or estimated these values to complete the model. For dust aggregates consisting of monodisperse,
spherical $\rm SiO_2$ particles with $0.75~\rm \mu m$ radius, our model is shown in Fig.
\ref{fig:model}. Each panel shows the mass-velocity parameter space for one of the eight
combinations of porous/compact and equal-sized/different-sized aggregate collisions. Green color
denotes the sticking (S1-S4) processes (mass gain), yellow color represents bouncing (B1-B2), and
red stands for the fragmentation cases F1-F3 (mass loss). The collision model does not only predict
the qualitative outcome of individual collisions between dust aggregates at different velocities,
but also the impact of the collision on the morphology of the collision product. This encompasses
(i) for sticking collisions, a recipe for the porosity of the new agglomerate, which can be higher
(e.g. process S1) or lower (e.g. process S3) than that of the colliding aggregates, (ii) for
bouncing collisions, the amount of compaction, (iii) for fragmenting collisions, the size
distribution of fragments and the mass of the largest fragment, and (iv) for all collision types
the amount of mass transfer between the aggregates. Before we get to the implications of our
dust-collision model for the evolution of dust aggregates in PPDs, we take another look at Fig.
\ref{fig:model} and try to find out which path the growth process can take. As stated in Sect.
\ref{sect:introduction}, the mean collision velocity of dust aggregates in PPDs increases with
increasing aggregate size between a few micrometers and one meter in size. Hence, the general
growth path must somehow proceed from the lower left to the upper right in Fig. \ref{fig:model}. In
all but one panels of Fig. \ref{fig:model}, this growth trajectory leads through bouncing or
fragmentation terrain. The only possible way towards larger dust aggregates can be found in the
`pP' panel, i.e. for collisions between porous dust aggregates of (very) different sizes. The
responsible growth process is then S2, in which the fluffy dust aggregates undergo some compaction
during the collision so that the contact area increases, which leads to sticking. As this process
is rather well characterized by the experimental results of \citet{langkowskietal2008}, we can have
some confidence that a dust evolution through this growth path can be realistically modelled.

\section{Dust evolution in protoplanetary disks}
\label{sect:results} 
\begin{figure}[htp]
   \centering
   \includegraphics[width=15.0cm, angle=0]{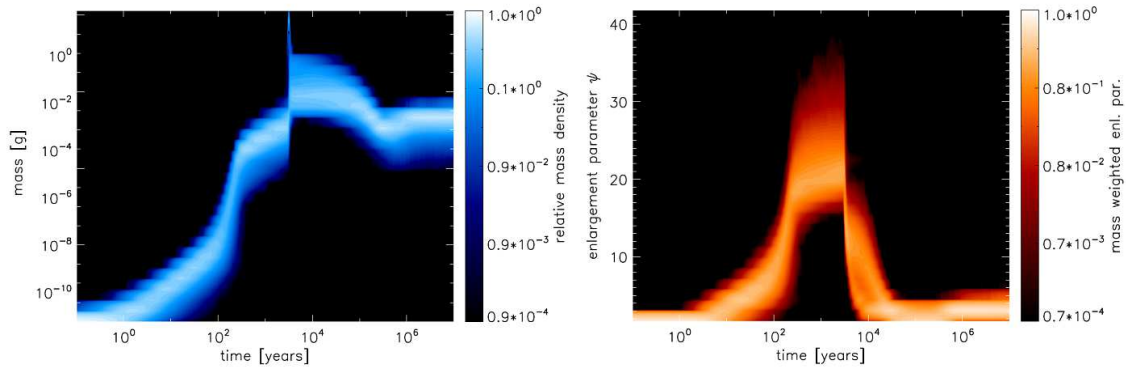}

   \caption{\label{fig:result}Results of the Monte-Carlo simulation of
   the dust evolution in PPDs (adapted from \citet{zsometal2010}).
   The underlying PPD model was a MMSN model at 1 AU and in the
   midplane with $\alpha = 10^{-4}$. The dust-aggregate collision model was taken from
   \citet{güttleretal2010}. The left graph shows the temporal
   evolution of the mass distribution of the aggregates, whereas
   the right graph denotes the temporal behavior of the enlargement parameter.}
   \end{figure}

With the `complete' dust-aggregate collision model described in Sect.
\ref{sect:model}, it is possible to predict the collisional evolution of protoplanetary dust
aggregates in any accretion-disk model. However, due to the complexity of the collision model,
solutions of rate equations are too multi-dimensional to allow sensible results. We chose the
Monte-Carlo method proposed by \citet{zsomdullemond2008} as the solver to the problem of dust
evolution in PPDs, using three different PPD models at three different $\alpha$ values each (see \citet{zsometal2010} for the details). As
a start and due to the considerable requirements in computational resources, we ran a local
simulation, addressing the dust-aggregate evolution at 1 AU from the central solar-type star and in
the midplane of the PPD. As relative velocities between the dust aggregates we considered Brownian
motion, radial drift and gas turbulence, the latter following the prescription of \citet{ormelcuzzi2007}. Of the three PPD models, the MMSN model resulted in the
largest achieved dust aggregates so that we will here discuss only the results of our simulations
of dust-aggregate evolution in the MMSN model. The main results of the study by
\citet{zsometal2010} can be seen in Fig. \ref{fig:result}), where in the left panel the temporal
evolution of the aggregate masses and in the right panel the corresponding enlargement parameters
are shown. The following main characteristics of the dust growth in PPDs must be emphasized:
\begin{itemize}
  \item From the nine physical processes, which constitute or collision model (see Fig.
  \ref{fig:model} and Sect. \ref{sect:model}), only four are entirely responsible for the grain
  evolution in the MMSN model. These are S1 (hit-and-stick growth), S2 (sticking through surface
  effects), B1 (bouncing with compaction), and B2 (bouncing with mass transfer).
  \item Grain growth is initially dominated by hit-and-stick collisions,
  in which fractal dust aggregates with $\sim 10^4$ monomer grains form
  within a few hundred years. After that, hit-and-stick (S1) and
  sticking with compaction (S2) dominate the growth for a few
  thousand years, forming high-porosity aggregates (with enlargement parameters of up to
  $\Psi = 15-35$) with masses of
  $\sim 10^{-3}$ g. A short period of runaway growth around
  $3 \times 10^3$ years then produces fluffy dust aggregates with masses
  of up to a few grams and with enlargement parameters of
  $\Psi \approx 10$. After that, the collisions among the dust
  aggregates are dominated by bouncing with compaction (B1), with
  minor contributions by bouncing with mass transfer (B2), which
  leads to a considerable reduction of the porosity. After a few $10^4$
  years, the dust aggregates are basically compact and cannot grow any
  further, due to a lack of smaller dust.
  \item When we take into account that compact dust aggregates can break
  up with a low probability, even at velocities smaller than the typical
  fragmentation threshold of $\sim 1~\rm m~s^{-1}$, as suggested by the experiments of
  \citet{weidlingetal2009}, the maximum aggregate mass decreases to
  $\sim 10^{-3}$ g. The timescale of the mass decrease (but not the
  final mass) depends on the fragmentation probability, which is not
  very well constrained by the present experiments.
  \item Aggregate growth stops at the `bouncing barrier', due to bouncing
  and not due to fragmentation, so that the maximum aggregate mass found
  in the model was not achieved by crossing the fragmentation barrier.
  In fact, fragmentation was never observed for the MMSN model.
  \item The dust-aggregate mass distribution in our model is always
  narrow so that small dust grains are depleted on short timescales.
  This might seem to contradict the observational evidence which
  supports retention of small dust particles over an extended period
  of time (e.g. \citet{furlanetal2005,kesslersilaccietal2006}), but we
  argue that these observations probe into the atmospheres of the PPDs
  (where the collision velocities are much higher so that grain growth
  can be inhibited) whereas our model was applied to the midplane.
  \item Increasing the collision velocities by applying stronger gas
  turbulence (with $\alpha$ values ranging from $\alpha = 10^{-5}$ to $\alpha = 10^{-3}$) provides some fragmentation but does not invoke the
  previously-proposed fragmentation-coagulation cycle \citep{teiserwurm2009},
  in which the fragmentation events provide sufficiently many small
  grains for a net growth to occur. The maximum aggregate mass increases sharply from $8 \times 10^{-2}$~g at $\alpha = 10^{-5}$ to 4~g at $\alpha = 10^{-4}$ and then slightly to 8~g at $\alpha = 10^{-3}$.
\end{itemize}

\section{Discussion and Outlook}
\label{sect:outlook} We now have, for the first time, a physical collision model for protoplanetary
dust aggregates, covering the full parameter space in aggregate mass and velocity and also dealing
with aspects of aggregate porosity and mass ratio of the colliding particles (Sect.
\ref{sect:model}). This model is based upon extensive laboratory work on the collision behavior of
dust aggregates. As we have seen in Sect. \ref{sect:results} (see also Fig. \ref{fig:result}),
Monte-Carlo type growth simulations based on this model predict a new `bouncing barrier' well below
the fragmentation threshold. That growth stops due to bouncing is a consequence of our collision
model and needs to be critically reviewed. The number of experimental investigations supporting the
dust-aggregate collision model is not small. However, the coverage of the huge (at least
four-dimensional) parameter space in our collision model is far from being complete. In fact, the
results of the Monte Carlo simulations show that the pathway of aggregate growth in PPD leads
through parameter terrain, which is not at all supported by experiments and which is subject to
large extrapolation uncertainties. The strength of a collaborative effort, bringing together
expertise from theory and modeling as well as results from laboratory investigations, is that we
can now start to investigate these unexplored regions with new and dedicated experiments. Two
examples of recently-developed laboratory experiments shall demonstrate this:
\begin{itemize}
  \item Collisions between very large and compact dust aggregates.\\
  Our dust-growth model predicts the formation of cm-sized compact
  aggregates (for $\alpha = 10^{-3}$ and $\alpha = 10^{-4}$) after about $10^4$ years (see
  Fig. \ref{fig:result}). The collision velocities of these aggregates
  in an MMSN model are $\sim 10 ~\rm cm~s^{-1}$. To investigate these
  collisions experimentally, we developed a laboratory mini drop tower, in
  which collisions between two dust aggregates with arbitrary
  composition in the size range between mm and dm and with collision
  velocities in the range $\sim 1 - 300\rm ~cm ~s^{-1}$ can be studied.
  First experiments with compact cm-sized dust aggregates showed
  that bouncing is the dominant collisional outcome in the relevant
  velocity regime around $10\rm ~cm ~s^{-1}$. We (Beitz, G\"uttler \& Blum, pers. comm.) found no sticking
  down to velocities below $1~\rm cm ~s^{-1}$ and fragmentation as
  the dominant process for velocities exceeding $\sim 100 ~\rm cm~s^{.1}$,
  in full agreement with the model and earlier work \citep{stewartleinhardt2009,setohetal2010}.
  \item The breakup probability of dust aggregates at velocities
  below the fragmentation limit.\\ The previous experiments by
  \citet{weidlingetal2009} had shown that with a probability of
  $\sim 10^{-4}$ dust aggregates can break up in collisions with
  velocities of $\sim 20~\rm cm~s^{-1}$, which is well below the
  fragmentation threshold of $100~\rm cm~s^{-1}$. As the final
  dust-aggregate size strongly depends on this breakup effect,
  we are interested in determining the breakup probability with a
  higher accuracy and as a function of collision velocity and aggregate properties. We
  therefore developed a new experimental setup in which a well-characterized
  single dust aggregate can undergo repeated collisions with a
  solid target at a well-defined impact velocity. First experiments
  have shown that the breakup probability of mm-sized dust aggregates
  increases considerably when the collision velocity is raised from
  $20\rm ~ cm~s^{-1}$ to $50\rm ~ cm~s^{-1}$ (Rott, G\"uttler \& Blum, pers. comm.).
  Once incorporated quantitatively in our collision model, this will surely
  change the predicted growth behavior of PPD dust aggregates.
  A detailed investigation is under way.
\end{itemize}
However, not every parameter combination which is predicted by the Monte-Carlo model is accessible
to experiments. Extremely low collision velocities and very high enlargement parameters are
basically impossible to achieve in the laboratory and the experimental aggregate-mass range is also
finite. A way out of this dilemma is a trustworthy collision model for dust aggregates of arbitrary
composition. Molecular dynamics models (see, e.g.,
\citet{dominiktielens1997,wadaetal2007,wadaetal2008,wadaetal2009}) are restricted to masses
$\stackrel{<}{\sim} 10^{-6}$ g so that a large part of the parameter space remains uncovered. We
have recently published a new Smooth Particle Hydrodynamics (SPH) model for dust-aggregate
collisions, based upon earlier work by \citet{sirono2004} and \citet{schaeferetal2007}, which might be helpful to fill the remaining parameter space with reliable predictions
for the collisional outcomes \citep{güttleretal2009,geretshauseretal2010}. Using dedicated
laboratory experiments, we were able to calibrate the SPH code for aggregates with an enlargement
parameter of $\Psi \approx 7$. Future investigations will concentrate on modelling the collision
behavior of dust aggregates with different enlargement parameters.

Our Monte-Carlo simulations indicated that there is no straightforward way for a direct growth of
km-sized planetesimals from small dust grains. It is obvious that the bouncing barrier is a stopper
for the previous fast growth of protoplanetary dust aggregates. However, these results depend on
the validity of both, our dust-aggregate collision model and the model for the collision velocities
of the dust particles. The former was developed for refractory silicates and might be more
favorable for sticking for icy materials, i.e. in the outer reaches of PPDs (laboratory work on the
potentially enhanced stickiness of $\rm \mu m$-sized ice particles is, however, still lacking). The
latter was derived by \citet{weidenschillingcuzzi1993} and \citet{ormelcuzzi2007} for a small
dust-to-gas ratio in the PPD. Due to the sedimentation of mm- to cm-sized particles, a
dust-dominated sub-disk can form in the midplane of the PPD (if turbulence is negligibly small,
i.e. in dead zones of otherwise MRI active PPDs), in which the collision velocities among the dust
aggregates can be suppressed. As the dust-aggregate collision model by \citet{güttleretal2010}
shows (see Fig. \ref{fig:model}), an unlimited growth of the protoplanetary dust is possible if
there is either a steady source of small, porous dust aggregates, or in collisions between compact
aggregates of very different sizes at reduced velocity. Our local aggregate growth model does not
predict a sufficiently wide size distribution to support this, but turbulence-driven mixtures of
dust aggregates from different regions could help to stimulate a further growth. Work on this is in
progress.

Alternatively, the collective gravitational effect of cm-sized dust aggregates in the dust sub-disk
can trigger the growth of macroscopic protoplanetary bodies
\citep{johansenetal2007,johansenetal2008,johansenetal2009}. Disk models with sufficiently high
global turbulence lead to transient high- and low-pressure zones, which result in a temporal and
local enhancement of the number density of dust aggregates. Such dust-enhanced regions rotate with
slightly higher velocity and forces the gas to a slightly faster rotation about the central star so
that the radial drift rate is locally decreased. This, in turn, means that dust aggregates from
outside the dust-enhanced regions can drift into the dust-enhanced clouds so that their mass
density is further increased. This `streaming instability' can ultimately lead to a gravitational
instability in the dust, as long as fragmentation events are suppressed \citep{johansenetal2008}.
The latest simulations predict the formation of planetesimals with sizes of 10-100 km within a few
orbital timescales \citep{johansenetal2009}, in agreement with the recent result from
\citet{morbidellietal2009} that asteroids started around this size. The current minimum size for
dust aggregates that can result in gravitational-unstable regions is a few cm, slightly above the
maximum size from our simulations. Future investigations will have to show whether this small gap
can be closed, particularly as our model shows that breakup of the cm-sized dust ultimately
decreases the aggregate sizes to millimeters. Alternatively, \citet{cuzzietal2010} developed a
model in which mm-sized particles (\citet{cuzzietal2010} assume chondrules but their physical
arguments also hold for equal-mass compact dust aggregates, which are the outcomes of our model)
concentrate in low-vorticity regions of a turbulent PPD, in which they can gravitationally interact
as an ensemble. \citet{cuzzietal2010} find that for PPD models with higher gas densities, higher
dust-to-gas ratios, and shallower radial density gradients than the MMSN model, bodies in the
100-km size range form on timescales compatible with direct age determinations of meteorites and
with the initial asteroid sizes derived by \citet{morbidellietal2009}.

Such gravitational instability scenarios also have the
advantage of avoiding the so-called `meter-size barrier'.
Meter-sized bodies in PPDs have radial drift timescales of the
order of 100 years, after which they are lost into the central
star. Any model that relies on the direct adhesional growth from
dust to planetesimals must pass this barrier within a time span
shorter than the lifetime of these bodies. This has been a problem
for the previous dust-aggregation model, but is naturally avoided
by gravitational instabilities.

It seems as if the two-stage planet-formation paradigm needs some refinement. Aggregation seems to
work nicely up to sizes of $\sim 1$ cm but is severely suppressed for larger dust-aggregates,
mainly owing to the bouncing barrier. However, due to the formation of a dust sub-disk, the
streaming instability can result in gravitationally-unstable regions in the dust, from which
asteroid-size bodies can collapse. The final stage in the formation of terrestrial planets is
accretion in mutual collisions, which has successfully been modelled in the past.

To assess the validity of such a three-stage planet-formation
model, more laboratory experiments, dust-aggregation simulations
in at least two spatial dimensions, and detailed investigations of
the collisional and gravitational evolution of dense
dust-aggregate ensembles are required.

\normalem
\begin{acknowledgements}
This work was funded by the German Space Agency (DLR) under grants Nos. 50WM0336, 50WM0636 and
50WM0936, and by the Deutsche Forschungsgemeinschaft (DFG) under grant No. Bl298/7-1. I thank
Carsten G\"uttler, Daniel Hei{\ss}elmann, Christopher Lammel, Ren\'{e} Weidling and Andras Zsom for
providing me with some of the Figures and Jeff Cuzzi for helpful comments on the manuscript.
\end{acknowledgements}

\label{lastpage}

\end{document}